\documentclass[aps,prd,twocolumn,floatfix,nofootinbib,showpacs,superscriptaddress]{revtex4}
\usepackage{graphicx,amsmath,amssymb,bm}
\usepackage{color}
\usepackage[utf8]{inputenc}
\definecolor{purple}{rgb}{0.5,0,0.5}
\definecolor{blue}{rgb}{0.0,0,0.9}
\usepackage[colorlinks=true, pdfstartview=FitV, citecolor= purple, linkcolor = blue,urlcolor=blue]{hyperref}

\begin{document}

\title{Interplay of dynamical and explicit chiral symmetry breaking effects on a quark}

\author{Fernando E. Serna}
\email{fernando.enrique@unesp.br}

\affiliation{Instituto Tecnol\'{o}gico de Aeron\'{a}utica, DCTA,
12228-900 S\~{a}o Jos\'{e} dos Campos, SP, Brazil}

\affiliation{Instituto de F\'{\i}sica Te\'{o}rica, Universidade Estadual Paulista, Rua Dr.~Bento Teobaldo Ferraz, 271 -- Bloco II, 01140-070 S\~{a}o Paulo, SP, Brazil}
\affiliation{Laborat\'orio de F\'isica Te\'orica e Computacional, Universidade Cruzeiro do Sul, Rua Galv\~ao Bueno, 868, 01506-000 S\~ao Paulo, S\~ao Paulo, Brazil}

\author{Chen Chen}
\email{chenchen@ift.unesp.br}
\affiliation{Instituto de F\'{\i}sica Te\'{o}rica, Universidade Estadual Paulista, Rua Dr.~Bento Teobaldo Ferraz, 271 -- Bloco II, 01140-070 S\~{a}o Paulo, SP, Brazil}
\author{Bruno~El-Bennich}
\email{bruno.bennich@cruzeirodosul.edu.br}
\affiliation{Instituto de F\'{\i}sica Te\'{o}rica, Universidade Estadual Paulista, Rua Dr.~Bento Teobaldo Ferraz, 271 -- Bloco II, 01140-070 S\~{a}o Paulo, SP, Brazil}
\affiliation{Laborat\'orio de F\'isica Te\'orica e Computacional, Universidade Cruzeiro do Sul, Rua Galv\~ao Bueno, 868, 01506-000 S\~ao Paulo, S\~ao Paulo, Brazil}

\begin{abstract}

The relative contributions of explicit and dynamical chiral symmetry breaking in QCD models of the quark-gap equation are studied in dependence of 
frequently employed ans\"atze for the dressed interaction and quark-gluon vertex. The explicit symmetry breaking contributions are defined by a constituent-quark 
sigma term whereas the combined effects of explicit and dynamical symmetry breaking are described by a Euclidean constituent-mass solution. We extend this
study of the gap equation to a quark-gluon vertex beyond the Abelian approximation complemented with numerical gluon- and ghost-dressing functions from 
lattice QCD.  We find that the ratio of the sigma term over the Euclidean mass is largely independent of nonperturbative interaction and vertex models for 
current-quark masses, $m_{u,d}(\mu) \leq m(\mu) \leq m_b(\mu)$, and equal contributions of explicit and dynamical chiral symmetry breaking
occur at $m(\mu) \approx 400$~MeV.  For massive solutions of the  gap equation with lattice propagators this value decreases to about 220~MeV.

\end{abstract}

\maketitle

\section{Introduction  \label{intro}}

Strong interactions are singularly characterized by a most effective mass-generating mechanism driven by dynamical chiral symmetry breaking (DCSB).
The scope and magnitude of the mass generation are unlike that observed in quantum electrodynamics, for example, and the origin of this chiral symmetry
breaking is thought to be intimately related with confinement~\cite{Bashir:2012fs}. Indeed, the emergence of a constituent quark-mass scale and the fact 
that DCSB contributes to nearly 98\% of visible mass has become a paradigm in contemporary hadron physics~\cite{Cloet:2013jya,Eichmann:2016yit}.
 
The impact of DCSB is evident in the light sector and plays an eminent role in describing why the nucleon's mass is about two orders of magnitude larger 
than that of its three bare constituents. For heavier quarks, starting with the strange quark, the effect of DCSB is gradually attenuated and the $b$-quark's 
constituent  mass is almost completely due to the Higgs mechanism~\cite{Ivanov:1998ms,Holl:2005st,ElBennich:2009vx,ElBennich:2012tp}. 

As a suitable measure for the effect of DCSB one can use the dimensionless ratio $\sigma_f / M_f^E$~\cite{Holl:2005st}, where $\sigma_f$ is the constituent 
quark's sigma term and $M_f^E$ is a Euclidean constituent-quark mass. This ratio quantifies the contribution of explicit chiral symmetry breaking (CSB)
to the dressed quark-mass function compared with the sum of both CSB and DCSB. It turns out that somewhere between the strange- and charm-quark 
mass the effects of CSB and DCSB are of the same order~\cite{Holl:2005st,ElBennich:2012tp}. Moreover, while the weak decay constants of light 
pseudoscalar and vector mesons increase with the light current-quark mass, they  level off somewhere between the  strange- and charm-quark mass 
and fall off for heavier quark masses as $f_M = 1/\surd{M_M}$~\cite{Bashir:2012fs,Maris:2005tt}. On the other hand, the weak decay constants of radially 
excited quarkonia can be shown to vanish in the chiral limit but though suppressed, their values increase again in a mass range somewhere between the 
$\bar ss$ and $\bar cc$ quarkonia~\cite{Rojas:2014aka,Mojica:2017tvh,El-Bennich:2017brb}. 

Clearly, dynamical effects on the dressed-mass function in  the intermediate range between these two current-quark mass scales have a substantial impact 
on hadronic observables; an analogue observation is that  SU$(4)_F$ flavor symmetry is badly broken compared with  
SU$(3)_F$~\cite{ElBennich:2010ha,ElBennich:2011py,El-Bennich:2016bno}.

In continuum approaches to Quantum Chromodynamics (QCD), such as the Dyson-Schwinger equation (DSE) for the quark, the strength of DCSB is governed 
by two ingredients in its integral kernel: the gluon dressing function~\cite{Aguilar:2008xm,Cyrol:2017ewj,Fischer:2008uz} and the dressed quark-gluon vertex
\cite{Davydychev:2000rt,Alkofer:2008tt,Rojas:2013tza,Rojas:2014tya,Qin:2013mta,Binosi:2014aea,Binosi:2016wcx,Aguilar:2010cn,Aguilar:2014lha,
Aguilar:2016lbe,Aguilar:2018epe,Aguilar:2018csq,Bashir:2011dp,Bermudez:2017bpx,Oliveira:2018fkj,Ball:1980ay,Ball:1980ax,Gao:2017tkg,Sultan:2018qpx}.
Failure to produce sufficient support result in a Wigner solution of the gap equation and thus any symmetry-preserving truncation must compensate for 
lacking interaction strength~\cite{Maris:1999nt}. The question arises how the pattern of DCSB and the relative effects between CSB and DCSB, for a given 
flavor, depend  on the simplifications applied to these kernels. It turns out that the contributions of CSB and DCSB to the constituent-quark mass are approximately 
similar halfway between the strange and charm current-quark masses  in the leading symmetry-preserving truncation of the quark's DSE 
and given functional form of the model interaction, namely the Maris-Tandy model~\cite{Maris:1999nt}. 

Including additional tensor structures of the dressed quark-gluon vertex is commonly compensated by readjusting the infrared-interaction strength of 
this model ``dressing" function; the functional form of the quark-mass function  as well as the position of associated complex-conjugate mass poles 
are consequently modified~\cite{El-Bennich:2016qmb}. While additional transverse vertex structures can be included in the DSE kernel without notable
computational efforts, this is not the case for the the integral kernel of antiquark-quark bound-state equations. Any Bethe-Salpeter kernel of the axialvector
vertex must satisfy an axialvector Ward-Takahashi identity and the truncation of the kernel must be consistent with that of the DSE. While this has been 
achieved to a certain degree (see the discussion in Ref.~\cite{Bashir:2012fs}), progress is still ongoing and necessary. We therefore limit ourselves
to explore the effects of DCSB in truncations of the DSE with increasing complexity and associated quark-$\sigma$ terms, and postpone more ambitious
calculations of hadronic $\sigma$ terms to future studies. 

We here contribute to gain additional insight in DCSB by studying the quark's DSE for different interaction and quark-gluon vertex ans\"atze, including the 
MT model, its more recent modification which reflects the results of modern DSE and lattice studies on the gluon propagator~\cite{Qin:2011dd}, the leading 
rainbow truncation, the Ball-Chiu vertex and transverse tensor structures of the vertex. It is found that  the behavior of the renormalization-point invariant ratio, 
$\sigma_f / M_f$, as a function of the current-quark mass  is nearly independent of the integral kernel formed by the convolution of the vertex- and gluon-dressing 
functions. On the other hand, for a given flavor and interaction tuned to reproduce light-hadron observables, the Euclidean mass, $M_f^E$, varies in a range of 
about 20--30\%. The extension of this numerical study to a gap equation with gluon and ghost propagators obtained with lattice QCD simulations mirrors 
the findings with model interactions.  

%%%%%%%%%%%%%%%%%%%%%%%%%%%%%%%%%%%%%%%%%%%%%%%%%%%%%%%%%%%%%%%%%%%%%%%%%%%%%%%%%%%

\section{Quark Dyson-Schwinger Equation}

\begin{figure}[t!]
\begin{center}
\includegraphics[scale=0.4]{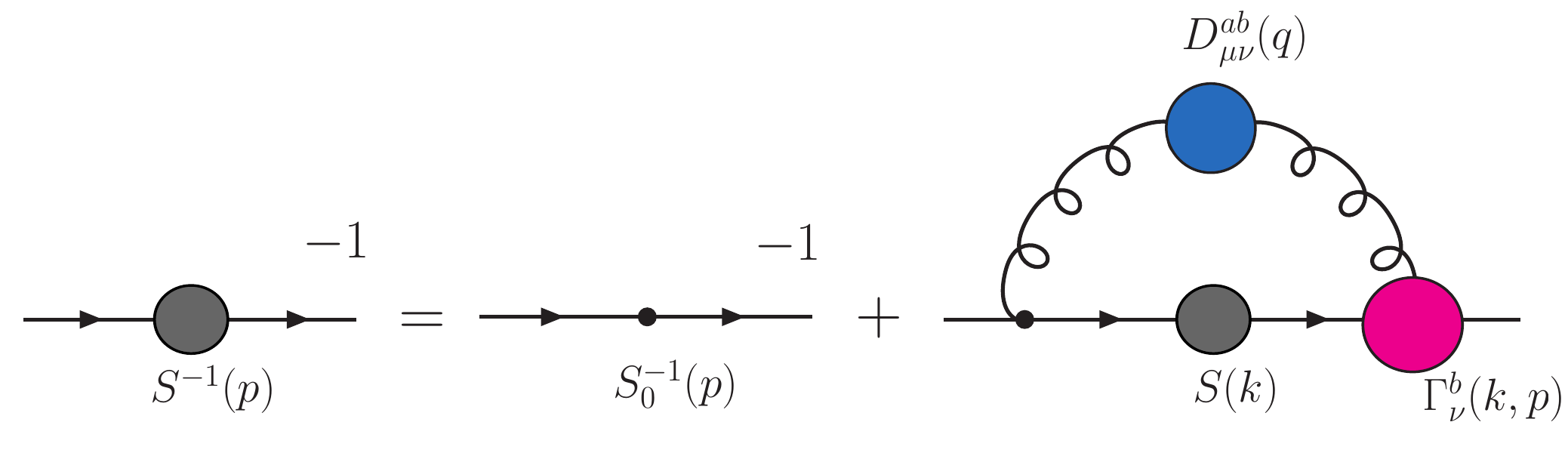}
\end{center}
\caption{Diagrammatic representation of the quark Dyson-Schwinger equation~\eqref{QuarkDSE}. Full circles denote fully-dressed propagators and vertices.}
\label{QDSE}
\end{figure}

The DSEs are the quantum equations of motion of a field theory and can be derived starting from the path integral formalism. As such, they are described by an infinite 
tower of coupled {\em exact\/} integral equations. In QCD, the quark fields obey a DSE~\cite{Maris:2003vk,Bashir:2004mu,Bashir:2012fs,Cloet:2013jya,Eichmann:2016yit}
for a given flavor, diagrammatically depicted in Fig.~\ref{QDSE},
%%%%%%%%%%%%%%%%%%%%%%%%%%%%%%%%%%%%%%%%%%%%%%%%%%%%%%%%%%%%%%%%%%%%%%%%%%%%%%%%%%%
\footnote{We employ throughout a Euclidean metric in our notation:  $\{\gamma_\mu,\gamma_\nu\} = 2\delta_{\mu\nu}$; $\gamma_\mu^\dagger = \gamma_\mu$; 
$\gamma_5= \gamma_4\gamma_1\gamma_2\gamma_3$, tr$[\gamma_4\gamma_\mu\gamma_\nu\gamma_\rho\gamma_\sigma]=-4\, \epsilon_{\mu\nu\rho\sigma}$; 
$\sigma_{\mu\nu}=(i/2)[\gamma_\mu,\gamma_\nu]$;  $a\cdot b = \sum_{i=1}^4 a_i b_i$; and $P_\mu$ timelike $\Rightarrow$ $P^2<0$.}
%%%%%%%%%%%%%%%%%%%%%%%%%%%%%%%%%%%%%%%%%%%%%%%%%%%%%%%%%%%%%%%%%%%%%%%%%%%%%%%%%%%
%
\begin{align}
S^{-1}_f (p)  & =   \, Z_2 \left (i\, \gamma \cdot  p + m^{\mathrm{bm}}_f \right ) \nonumber \\
                & +   \, Z_1\, g^2\!  \int^\Lambda_k \!  D^{ab}_{\mu\nu} (q) \frac{\lambda^a}{2} \gamma_\mu\, S_f(k) \,\Gamma^b_\nu (k,p) \ ,
\label{QuarkDSE}
\end{align}
where $q=k-p$, $Z_1(\mu,\Lambda )$ and $Z_2(\mu,\Lambda )$ are the vertex and quark wave-function renormalization constants, respectively,  and $\mu$ is the renormalization
point. Infinite radiative gluon corrections yield the quark self energy which modify the current quark bare mass, $m^{\mathrm{bm}}_f(\Lambda)$,  and where the integral is 
over the dressed gluon propagator, $D_{\mu\nu}(q)$, and the dressed quark-gluon vertex, $\Gamma^a_\mu (k,p) = \frac{1}{2}\lambda^a \,\Gamma_\mu (k,p)  $; 
the color SU(3) matrices, $\lambda^a$, are in the fundamental representation. In the integral, the abbreviation $\int_k^\Lambda\equiv \int^\Lambda d^4k/(2\pi)^4$ represents 
a Poincar\'e-invariant regularization with the regularization-mass scale $\Lambda$.  We work in Landau gauge, where the gluon propagator is purely transversal,
\begin{equation}
   D^{ab}_{\mu\nu} (q) =  \delta^{ab} \left( \delta_{\mu\nu} - \frac{q_\mu q_\nu}{q^2} \right) \frac{\Delta ( q^2) }{q^2} \ ,
\end{equation}   
which defines the gluon-dressing function $\Delta (q^2)$. 

The solutions to the gap equation~(\ref{QuarkDSE})  for spacelike momenta, $p^2>0$, can be decomposed into a vector and scalar piece, 
\begin{equation}
   S_f (p) =  \left [ i\, \gamma\cdot p\ A_f(p^2) +\mathbb{I}_D\,  B_f(p^2) \right ]^{-1} \  ,
\label{sigmaSV}
\end{equation}
and the renormalization condition,
\begin{equation}
\left. Z_f (p^2) = 1/A_f (p^2)  \right |_{p^2 = \mu^2} = 1 \  ,
\label{EQ:Amu_ren}
\end{equation}
is imposed. Typically, in conjunction with the MT model discussed below, $A_f (p^2)$ is renormalized at a large spacelike momentum, $\mu = 19$~GeV $ \gg \Lambda_\mathrm{QCD}$. 
More recent numerical results of the quark propagator's dressing functions from lattice QCD also allow for a much lower renormalization point $A_f (\mu \simeq 2~\mathrm{GeV}) = 1$.
The mass function,  $M_f(p^2)=B_f(p^2\!, \mu^2)/A_f(p^2\!, \mu^2)$, is independent of  $\mu$. The scalar function $B_f(p^2)$ is determined with another 
renormalization  condition,
\begin{equation}
\left.  S^{-1}_f(p) \right |_{p^2=\mu^2}  = \  i\  \gamma\cdot p \ + m_f(\mu )\, \mathbb{I}_D  \ ,
\label{massmu_ren}
\end{equation}
where $m_f(\mu )$ is the renormalized running quark mass:
\begin{equation}
\label{mzeta} 
   Z_m^f  (\mu,\Lambda )\, m_f(\mu)  =  m_f^{\rm bm} (\Lambda) \  .
\end{equation}
Here, $Z_m^f(\mu,\Lambda ) =  Z_4^f(\mu,\Lambda )/Z_2^f (\mu,\Lambda )$ is the flavor dependent mass-renormalization constant and $Z_4^f(\mu,\Lambda )$ is the renormalization 
constant associated with the Lagrangian's mass term; $Z_2$ and $Z_4$ are fixed by the renormalization conditions in Eqs.~\eqref{EQ:Amu_ren} and \eqref{massmu_ren}.
Notably, $m_f(\mu )$ is  the quark-mass function evaluated at a particular deep spacelike point, $p^2=\mu^2$, which makes contact with perturbative QCD:
\begin{equation}
  m_f(\mu)  = M_f(\mu )\,.
\end{equation}
Finally, we remark that the renormalization-group invariant current-quark mass can be inferred from,
\begin{equation}
  \hat m_f =  \lim_{p^2 \to \infty} M_f (p^2) \left [ \frac{1}{2} \ln\left ( \frac{p^2}{\Lambda_\mathrm{QCD}^2}\right ) \right ]^{\gamma_m} \  ,
 \label{invariantmass}
\end{equation}
where $\gamma_m$ is the anomalous mass dimension.

%%%%%%%%%%%%%%%%%%%%%%%%%%%%%%%%%%%%%%%%%%%%%%%%%%%%%%%%%%%%%%%%%%%%%%%%%%%%%%%%%%%

\subsection{Quark-Gluon Vertex}
\label{quarkgluon}

Due to asymptotic freedom, the behavior of the kernel at large momenta is known in perturbation theory in the domain $q^2 \simeq p^2 \simeq k^2 \gtrsim 2$~GeV$^2$
from which one can derive a sensible model for realistic DSE calculations~\cite{Binosi:2014aea} given by, 
\begin{equation}
\hspace*{-1mm}
   Z_1\, g^2  D_{\mu\nu} (q) \,  \Gamma_\mu (k, p)  =     Z_2\, q^2 \mathcal{G} (q^2)  D_{\mu\nu}^\mathrm{free} (q) \, \Gamma_\mu (k,p)\,  ,
\label{DSEtrunc}  
\end{equation}
where $ D_{\mu\nu}^\mathrm{free}(q) = \big ( \delta_{\mu\nu} -  q_\mu q_\nu/q^2 \big )/q^2$ is the free gluon propagator. In Eq.~\eqref{DSEtrunc}, the Abelianized 
Ward-Green-Takahashi identity  (WGTI),
\begin{equation}
  q_\mu i \Gamma_\mu (k,p) = S^{-1} (k) - S^{-1} (p) \ ,
  \label{WTIequation}
\end{equation}
has been enforced, $Z_1=Z_2$,  which at one loop corresponds to neglecting the contributions of the three-gluon vertex 
to $\Gamma_\mu (k, p)$. Formally, this is  equivalent to setting the renormalization constants for the ghost-gluon vertex and ghost wave function equal: 
$\tilde Z_1 = \tilde Z_3$. 

In the leading truncation we employ the ansatz,
\begin{equation}
  \Gamma_\mu (k, p) =  Z_2 \, \gamma_\mu  \  ,
\label{RLvertex}  
\end{equation}
where an additional factor $Z_2$ is included~\cite{Bloch:2002eq} to ensure multiplicative renormalizability of Eq.~\eqref{QuarkDSE} and thus the renormalization-point independence 
of $M(p^2)$. When the Abelian approximation, $Z_1= Z_2$, is used along with the {\em rainbow-ladder} (RL) truncation truncation, $Z_1=1$, it preserves the one-loop anomalous 
dimension of  $M(p^2)$~\cite{Maris:1997tm}. Therefore, to make contact with early studies on the quark DSE, we also absorb the renormalization constants $Z_2^2$ from 
Eqs.~\eqref{DSEtrunc} and \eqref{RLvertex} into the function $\mathcal{G} (q^2)$ in case of the MT  model~\eqref{MT}, and only in that case, which effectively describes the effects 
of both, the gluon and the vertex dressing.

To go beyond this approximation, we treat the case  of the Ball-Chiu (BC) ansatz~\cite{Ball:1980ay,Ball:1980ax} which satisfies Eq.~\eqref{WTIequation} by construction,
\begin{align}
  \Gamma_\mu^\mathrm{BC} (k, p) & =   \Sigma_A(k^2,p^2)\gamma_\mu +\Delta_A(k^2,p^2) \gamma \cdot\! (k +  p) (k+p)_\mu  \nonumber  \\
                                                        & -   i \, \Delta_B(k^2,p^2) (k+p)_\mu \, \mathbb{I}_D\, ,
\label{BallChiu}                                                        
\end{align}
with the compact definitions,
\begin{eqnarray*}
    \Sigma_\phi(k^2,p^2)  & = &  \frac{\phi(k^2)+ \phi(p^2)}{2}, \\ 
    \Delta_\phi(k^2,p^2)    & = &  \frac{\phi(k^2)-\phi(p^2)}{k^2-p^2}\, ,
\end{eqnarray*}
and $\phi (k^2) = A  (k^2), B (k^2)$.
The vertex in Eq.~\eqref{BallChiu} clearly implies a flavor dependence via the vector and scalar functions $A_f(p^2)$ and $B_f(p^2)$.  

The WGTI only constrains the Ball-Chiu components of $\Gamma_\mu (k, p)$ but extensive studies in perturbation theory have also shed light on the functional dependence of the 
transverse vertex components on $A(p^2)$ and $B(p^2)$~\cite{Bashir:2011dp,Bermudez:2017bpx} in certain kinematic limits. We here use an ansatz~\cite{Chang:2012cc} which 
models the anomalous chromomagnetic moment,
\begin{align}
\Gamma^\mathrm{T}_\mu (k,p)  & =   \tfrac{1}{2}\left[(k+p)^{\rm T}_\mu\, \gamma \cdot q   +\gamma^{\rm T}_\mu  \sigma_{\nu\rho}(k+p)_\nu\,  q_\rho\, \right ]  \nonumber    \\ 
                                                    &     \times \,  \frac{\eta\,\Delta_B(k^2,p^2)}{\mathcal{M}(k^2,p^2)}  +  \eta \, \Delta_B(k^2,p^2)\, \sigma_{\mu\nu} q_\nu  \ ,
\label{transverse}                                                    
\end{align}
with $\ell^T_\mu : = T_{\mu\nu} \ell_\nu$, $T_{\mu\nu} = \delta_{\mu\nu} - q_\mu q_\nu/q^2$,  $\eta=0.325$ and introducing the function:
\begin{equation*}
  \mathcal{M}(k^2,p^2)  =   \frac{k^2+M^2(k^2)+p^2+M^2(p^2)}{2 \left [ M(k^2)+M(p^2) \right ]  } \ .
\end{equation*}
Adding these transverse components to the Ball-Chiu vertex the nonperturbative quark-gluon vertex  becomes the linear sum,
\begin{equation}
   \Gamma_\mu (k,p)  =     \Gamma_\mu^\mathrm{BC} (k, p)  +   \Gamma^\mathrm{T}_\mu (k,p) \ . 
 \label{fullvertex}  
\end{equation}

Independent of perturbative results, the transverse dressing functions can be derived from Lorentz symmetries using a set of transverse WGTIs~\cite{Kondo:1996xn,Pennington:2005mw,He:2006my} 
and their functional form in the Abelian case can be found in Refs.~\cite{Qin:2013mta,Aguilar:2014lha}. In QCD, on the other hand, the fermion-gauge vertex satisfies a Slavnov-Taylor 
identity (STI)~\cite{Slavnov:1972fg,Taylor:1971ff}  which also leaves the transverse component undetermined. The studies of the Abelian transverse vertex identities were generalized to
transverse STIs which lead to expressions that depend on the scalar- and vector-quark functions but also on the ghost dressing function, the quark-ghost scattering amplitude and on 
a nontrivial, nonlocal four-point function which is a consequence of gauge invariance~\cite{He:2009sj}. The latter term involves a Wilson line in QED and QCD and can be parametrized 
most generally by four tensor structures and corresponding form factors; similarly the most general quark-ghost scattering kernel consists of four matrix-valued amplitudes which can be 
computed  within a nonperturbative {\em dressed} propagator model~\cite{Rojas:2014aka,Aguilar:2010cn,Aguilar:2016lbe,Aguilar:2018epe,Aguilar:2018csq}.

Expressions for the transverse vertex function derived from transverse STIs and constrained by pQCD~\cite{Bashir:2011dp} that satisfy multiplicative renormalizability will be presented 
elsewhere~\cite{Ahmed2018}. A simplified, minimal form of this novel vertex can be expressed by,
\begin{eqnarray}
\Gamma_\mu^\mathrm{STI} (k,p)  & = &  G(q^2) X_0(q^2)  \Big [ \Sigma_A(k^2,p^2) \gamma_\mu         \nonumber \\
                                                       & + &  \Delta_A(k^2,p^2)  \,  \gamma \cdot (k +  p) (k+p)_\mu     \nonumber \\
                                                       & - &  i \Delta_B (k^2,p^2)   (k+p)_\mu \, \mathbb{I}_D         \nonumber \\
                                                       & + &  \frac{\Delta_A(k^2,p^2) }{2}  \left (q^2\gamma_\mu - \gamma \cdot q\,  q_\mu \right )   \nonumber \\
                                                       & - &   \Delta_B(k^2,p^2) \,    \sigma_{\mu \nu} \gamma_\nu  \Big ]  \ ,
%                                                       & + &  \Delta_A(k^2,p^2) \big ( -i \gamma_{\mu} \, p^{\nu} k^{\rho} \sigma_{\nu \, \rho}   \nonumber \\
%                                                       & - &   p_{\mu} \, \gamma \cdot k + k_{\mu}\, \gamma \cdot p \big )  \Big ]  \ ,
\label{STIvertex} 
\end{eqnarray}
where $G(q^2)$ is the ghost dressing function and $X_0(q^2)$ the leading form factor of the quark-ghost kernel~\cite{Rojas:2013tza,Aguilar:2010cn}. Additional form factors
for arbitrary momenta were obtained in Ref.~\cite{Aguilar:2018epe} and can readily be included. For the present purpose their contributions can be neglected and we also set
$X_0(q^2) = 1$~\cite{Rojas:2013tza,Rojas:2014tya}.

%%%%%%%%%%%%%%%%%%%%%%%%%%%%%%%%%%%%%%%%%%%%%%%%%%%%%%%%%%%%%%%%%%%%%%%%%%%%%%%%%%%

\subsection{Gluon Interaction Models and Lattice QCD Dressing Functions}
\label{gluon}

An interaction ansatz for $\mathcal{G} (q^2)$ that has proven its merits in meson and baryon phenomenology is the MT model~\cite{Maris:1999nt} and can 
be decomposed as, 
\begin{equation}
    \mathcal{G} (q^2) = q^2\mathcal{G}_\mathrm{ IR}(q^2) +  4\pi \tilde\alpha_\mathrm{PT} (q^2)  \ ,
\label{IR+UV}    
\end{equation}
where $\tilde\alpha_\mathrm{PT} (q^2) $ is a monotonically decreasing and regular continuation of the perturbative strong coupling in QCD and $\mathcal{G}_\mathrm{ IR}(q^2)$
is an ansatz for the interaction in the infrared domain of squared momenta. $\mathcal{G}_\mathrm{ IR}(q^2)$ is strongly suppressed for $q^2 \gtrsim 2$~GeV$^2$ where 
$\tilde\alpha_\mathrm{PT} (q^2) $ dominates.

In all instances we use,
\begin{equation}
     4\pi \tilde\alpha_\mathrm{PT}(q^2) =  \frac{8\pi^2 q^2 \gamma_m \mathcal{F}(q^2)}{ \ln \left [  \tau +\left (1 + q^2/\Lambda^2_{\rm QCD} \right )^{\!2} \right ] }  \ ,
\label{alphaPT}     
\end{equation}
with $\gamma_m=12/(33-2N_f)$, $N_f = 4$, $\Lambda_{\rm QCD}=0.234$GeV, $\tau=e^2-1$ and $\mathcal{F}(q^2)=[1-\exp(q^2/4m^2_t)]/q^2$, $m_t=0.5$~GeV.
This functional form preserves the one-loop renormalization-group behavior of QCD in the gap equation. The low-momentum range of the MT model is 
described by a Gaussian-type support that vanishes in the infrared, 
\begin{equation}
  \mathcal{G}^\mathrm{MT}_\mathrm{IR} (q^2)  =   \frac{4\pi^2}{\omega^6}q^2 De^{-q^2/\omega^2} \ .
\label{MT}
\end{equation}
More recently, though, a modified version of this function was proposed~\cite{Qin:2011dd} which deliberately avoids the $q^2$-factor and therefore leads to an infrared massive 
and finite interaction:
\begin{equation}
  \mathcal{G}^\mathrm{QC}_\mathrm{IR} (q^2)  =   \frac{8\pi^2}{\omega^4} De^{-q^2/\omega^2} \ .
\label{QC}
\end{equation}
We stress that neither models, Eqs.~\eqref{MT} and \eqref{QC},  in conjunction with  Eqs.~\eqref{IR+UV}  and \eqref{alphaPT} can be expressed via a non-negative spectral density. 
It is also a feature of these interactions that they are virtually insensitive to variations of $0.4\leq \omega \leq 0.6$ so long as the product $D\omega$ remains constant. It is crucial, though, 
that the form and parametrization of these models provide enough strength to realize sufficient DCSB. We here use both interaction ans\"atze in  studying  the interplay of effects of 
CSB and DCSB for different quark flavors in Section~\ref{Results}.

Along with the vertex ansatz in Eq.~\eqref{STIvertex} derived from longitudinal and transverse STIs, we make use of Pad\'e approximations~\cite{Bashir:2013zha,Binosi:2016xxu} 
for unquenched ($N_f $ = 2+1+1) lattice-regularized ghost- and gluon-dressing functions, $G^\mathrm{Latt.} (q^2)$ and $\Delta^\mathrm{Latt.}  (q^2)$ respectively~\cite{Ayala:2012pb}. 
The integral kernel in  Eq.~\eqref{DSEtrunc} thus becomes,
\begin{align}
  Z_1\, g^2  D_{\mu\nu} (q) \,  \Gamma_\nu (k, p) =  &  \nonumber\\[0.2true cm]
  Z_2 \, 4\pi \alpha_s \, q^2 \Delta^\mathrm{\! Latt.} & (q^2)     D_{\mu\nu}^\mathrm{free} (q) \, \Gamma_\nu^\mathrm{STI}  (k,p)  \,  .
\label{STIkernel}  
\end{align}
%

%%%%%%%%%%%%%%%%%%%%%%%%%%%%%%%%%%%%%%%%%%%%%%%%%%%%%%%%%%%%%%%%%%%%%%%%%%%%%%%%%%%%%%%%%%%%%%%%%%%%%%%%%%%%%%%%%%%%%%%%%%%%%%%%%%%%%%%%%%%%%%%%%%%%%%%%%%%%%%%%%%%%%

\subsection{ The Quark Sigma Term and Constituent Quark Mass}

A convenient parameter to study the effect of DCSB is the renormalization-point invariant ratio,
\begin{equation}
   \zeta := \frac{\sigma_f}{M^E_f}  \ ,
\label{zetadef}   
\end{equation}
where $\sigma_f$ is the constituent-quark  sigma term and $M^E_f$ is the Euclidean constituent mass. In analogy with the nucleon's sigma term, one defines a measure 
of the contribution from CSB to the constituent quark mass by  $\sigma_f : = m_f(\mu) \langle Q | \bar q_f q_f | Q \rangle $ and using the Hellmann-Feynman theorem
\cite{Hellmann:1933,Feynman:1939zza}  this scalar matrix element can be directly related to the constituent-quark mass, 
\begin{equation}
   \sigma_f = m_f(\mu)\, \frac{\partial M^E_f}{\partial m_f(\mu)} \ ,
\end{equation}
where $M^E_f$ is the the Euclidean  mass functions solution~\cite{Bashir:2012fs}: 
\begin{equation}
    (M^E_f)^2:  =  \left \{ p^2|p^2=M^2(p^2)\right  \}  \ . 
\label{euclidmass}    
\end{equation}  
Since at the quark level, $M^E_f$ contains both, the CSB and DCSB contributions to the quark's mass, the ratio  $\zeta$ is zero in the chiral limit and increases with larger 
current-quark mass: $0 \leq \zeta \leq 1$. The case $\zeta \simeq 1$ is expected for the top-quark mass. 

It should be mentioned that the definition of a constituent-quark sigma term independent of its hadronic environment is problematic due to interactions with bystander quarks 
and chiral corrections are important for light quarks; indeed, the definition of $\sigma_f$ ought to depend on the hadron's properties~\cite{Thomas:2000fa,Leinweber:2004tc}.
Naturally, the effects of DCSB are more comprehensively studied with hadronic $\sigma$ terms, where they measure the contribution of non-vanishing current 
quark masses to the nucleon mass. 

Here, we merely want to investigate whether different DSE kernels actually affect the relative contributions of DCSB to CSB to the quark mass, as it is 
well known that the running and functional behavior of the mass function depend on the dressed-vertex ansatz; our solutions for $M(p^2)$ for all vertex
ans\"atze confirm this. To this end, the definition of Eq.~\eqref{zetadef}   provides a  renormalization-point independent measure. In particular, as our results indicate, 
the charm quark is far from being a heavy quark and in any of the truncations we consider here, the DCSB contribution to its mass is about 40\%. Thus, the frequent 
use of a constituent charm quark in calculations of charmed masses and form factors is inadequate. Moreover, the relative contributions of CSB and DCSB are
fairly independent of the DSE kernel's truncation.

 %%%%%%%%%%%%%%%%%%%%%%%%%%%%%%%%%%%%%%%%%%%%%%%%%%%%%%%%%%%%%%%%%%%%%%%%%%%%%%%%%%%

\begin{table}[t!]
\caption{Interaction parameters employed in Maris-Tandy (MT) and Qin-Chang (QC) models, Eqs.~\eqref{MT} and \eqref{QC} respectively, in combination with the 
three vertex ans\"atze  described in Section~\ref{quarkgluon}. In case of the MT model we absorb the renormalization constant $Z_2$ in the interaction 
$\mathcal{G}^\mathrm{MT}(q^2)$ but include it in the vertex when using the QC model; see Eq.~\eqref{DSEtrunc}.}

\def\arraystretch{1.5}
\centering\vspace*{1mm}
\begin{tabular*}
{\hsize}
{
c@{\extracolsep{0ptplus1fil}}
c@{\extracolsep{0ptplus1fil}}
c@{\extracolsep{0ptplus1fil}}
}
\hline\hline
 \textrm{ Model Parameters }  & $(\omega D)^{1/3}$ [GeV]   &   $\omega$ [GeV]  \\
\hline\hline
$\text{MT+RL}$ & 0.72& 0.40   \\
$\text{QC+RL}$ & 0.80&0.50   \\
$\text{MT+BC}$ & 0.65& 0.50 \\
$\text{QC+BC}$ &0.65 &0.50 \\
$\text{MT+(BC+T)}$ &0.55&0.50  \\
$\text{QC+(BC+T)}$ &0.52&0.50  \\\hline\hline
\end{tabular*} 
\label{Table:Results}
\end{table}

%%%%%%%%%%%%%%%%%%%%%%%%%%%%%%%%%%%%%%%%%%%%%%%%%%%%%%%%%%%%%%%%%%%%%%%%%%%%%%%%%%%
%%%%%%%%%%%%%%%%%%%%%%%%%%%%%%%%%%%%%%%%%%%%%%%%%%%%%%%%%%%%%%%%%%%%%%%%%%%%%%%%%%%

\section{Numerical Results: DCSB and CSB interplay and interaction kernels}
\label{Results}

The quark-gap equation~\eqref{QuarkDSE} is numerically solved for the three quark-gluon vertices and dressed interaction models detailed in Sections~\ref{quarkgluon} and 
\ref{gluon}, where the model parameters $\omega$ and $D$ are chosen such that for a given interaction kernel the experimental light-hadron mass spectrum and 
weak decay constants are reproduced. We collect them in Table~\ref{Table:Results} and use the same parameters for all quark flavors. 

The renormalization conditions~\eqref{EQ:Amu_ren} and \eqref{massmu_ren} are imposed at $\mu=19$~GeV for the vertices in Eqs.~\eqref{RLvertex}, \eqref{BallChiu}
and \eqref{fullvertex} for both the MT and QC interactions. The bare vertex is multiplied by a renormalization constant $Z_2$ as in Eq.~\eqref{RLvertex}, which introduces 
a renormalization factor $Z_2^2$ on the right-hand-side of Eq.~\eqref{DSEtrunc}. In case of the MT interaction we always absorb $Z_2^2$ in  the interaction function 
$\mathcal{G}(q^2)$ whereas it is explicitly maintained for the QC interaction; this leads to linear and nonlinear renormalization conditions, respectively~\cite{Hilger:2017jti}. 

\begin{figure}[t]
\begin{center}
\vspace*{-4mm}
\hspace*{-4mm}
  \includegraphics[scale=0.6]{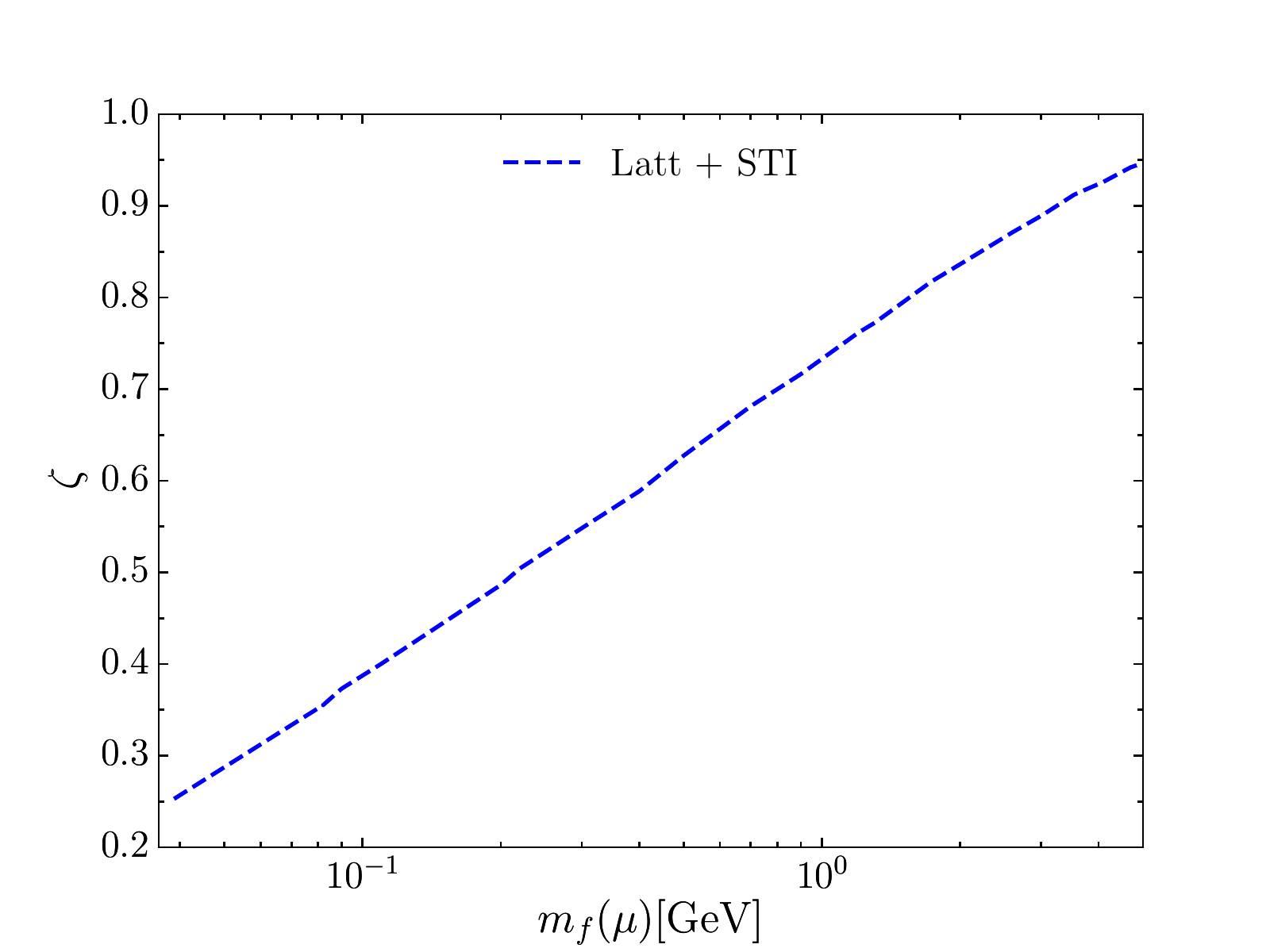}
\end{center}
\caption{The renormalization-point invariant ratio $\zeta$ as a function of $m_f(\mu)$ obtained with the interaction kernel Eq.~\eqref{STIkernel}. }
\label{zetalattice}
\end{figure}

The STI vertex~\eqref{STIvertex} is treated somewhat differently: as indicated in Eq.~\eqref{STIkernel}, a linear renormalization condition is imposed but we choose the
scale $\mu=4.3$~GeV at which the unquenched  ($N_f $ = 2+1+1) dressing functions $G^\mathrm{Latt.} (q^2)$ and $\Delta^\mathrm{Latt.} (q^2)$ were renormalized
\cite{Ayala:2012pb}.  Likewise, we employ the light and heavy renormalized quark masses of that same reference: $m_{u,d}(2~\mathrm{GeV} ) \simeq 40$~MeV,  
$m_s (2~\mathrm{GeV} ) = 95$~MeV, $m_c (2~\mathrm{GeV} ) =1.51$~GeV evolved to the scale $\mu=4.3$~GeV and for the beauty quark we choose 
$m_{b}(4.3~\mathrm{GeV})=M_b(4.3~\mathrm{GeV})=4.697$~GeV obtained from the solution of the quark DSE~\eqref{QuarkDSE} with the interaction produced by 
Eqs.~\eqref{IR+UV}, \eqref{alphaPT} and~\eqref{QC} and the  BC +T vertex~\eqref{fullvertex}. For comparison, we also solve the same DSE with quenched $(N_f=0)$
and partially quenched $(N_f=2)$ gluon and ghost propagators which are less suppressed than the unquenched ones, however this is compensated by a decrease 
of the corresponding value of $\alpha_s^{N_f}\! (\mu)$~\cite{Ayala:2012pb} and the evolution of $\zeta$ in Fig.~\ref{zetalattice} is fairly independent of $N_f$.

%%%%%%%%%%%%%%%%%%%%%%%%%%%%%%%%%%%%%%%%%%%%%%%%%%%%%%%%%%%%%%%%%%%%%%%%%%%%%%%%%%%

%%%%%%%%%%%%%%%%%%%%%%%%%%%%%%%%%%%%%%%%%%%%%%%%%%%%%%%%%%%%%%%%%%%%%%%%%%%%%%%%%%%

\begin{table}[t!]
\caption{ Euclidean quark masses $M^E_f$ (GeV) as defined in Eq.~\eqref{euclidmass} with the model parameters of Table~\ref{Table:Results}. We assume isospin 
symmetry for the two lightest quarks.}
\def\arraystretch{1.5}
\centering\vspace*{1mm}
\begin{tabular*}%{|c|c|c|c|c|c|c|}\hline
{\hsize}
{
c@{\extracolsep{0ptplus1fil}}
c@{\extracolsep{0ptplus1fil}}
c@{\extracolsep{0ptplus1fil}}
c@{\extracolsep{0ptplus1fil}}
c@{\extracolsep{0ptplus1fil}}
}
\hline\hline
$f$ & $u,d$ & $s$ & $c$&$b$\\
\hline\hline 
$m_f(\mu)$ & 0.0037& 0.082 & 0.970&4.100 \\ \hline \hline  
$(M^E_f)^{\text{MT+RL}}$ & 0.403& 0.555 & 1.566  & 4.682 \\ 
$(M^E_f)^{\text{QC+RL}}$ & 0.408& 0.563 & 1.576 &  4.701  \\
$(M^E_f)^{\text{MT+BC}}$ & 0.385& 0.512 & 1.505  &  4.648  \\ 
$(M^E_f)^{\text{QC+BC}}$ & 0.381& 0.495 & 1.495  &  4.664  \\ 
$(M^E_f)^{\text{MT+(BC+T)}}$ & 0.387& 0.533 & 1.549  &  4.693  \\
$(M^E_f)^{\text{QC+(BC+T)}}$ & 0.390& 0.514 & 1.530  &  4.687  \\ \hline\hline
\end{tabular*} 
\label{EQM}
\end{table}

%%%%%%%%%%%%%%%%%%%%%%%%%%%%%%%%%%%%%%%%%%%%%%%%%%%%%%%%%%%%%%%%%%%%%%%%%%%%%%%%%%%%%%%%%%%%%

\begin{figure}[t!]
\vspace*{-3mm}
\hspace*{-3mm}
  \includegraphics[scale=0.6]{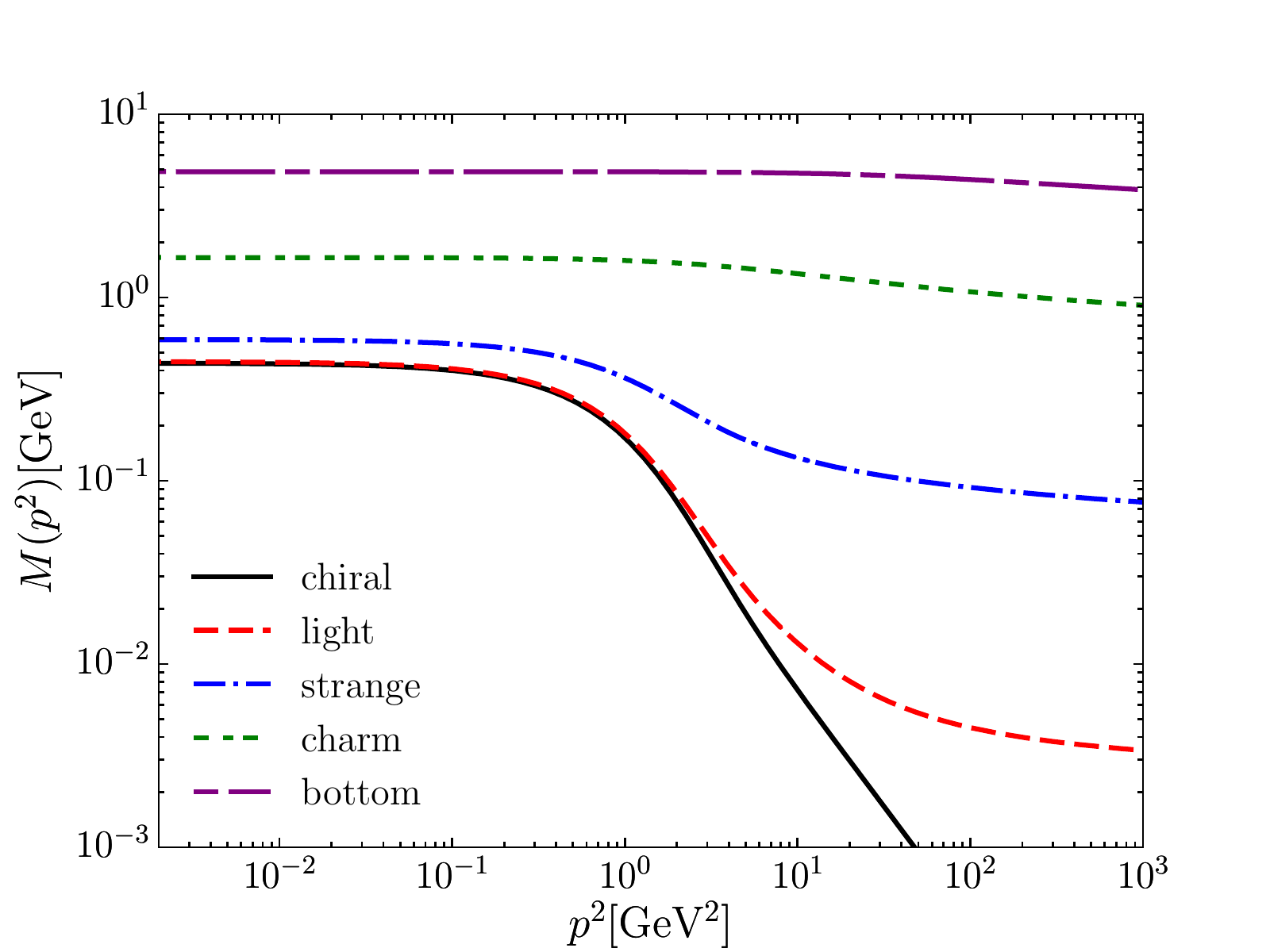}
\caption{The running mass, $M_f(p^2)$, generated by the interaction produced by Eqs.~\eqref{IR+UV}, \eqref{alphaPT} and~\eqref{QC} with  $(\omega D)^{1/3}= 0.52$~GeV, 
$\omega = 0.50$, and the dressed quark-gluon vertex of Eqs.~\eqref{BallChiu} to \eqref{fullvertex} with $\eta=0.325$. This leads to $M_{\hat m_f =0}(0)=0.434$~GeV in the 
chiral limit,  $M_{u,d}(0)=0.447$~GeV,  $M_s(0)=0.590$~GeV, $M_c(0)=1.654$~GeV and $M_b(0)=4.865$~GeV. }
\label{Massf}
\end{figure}
%%%%%%%%%%%%%%%%%%%%%%%%%%%%%%%%%%%%%%%%%%%%%%%%%%%%%%%%%%%%%%%%%%%%%%%%%%%%%%%%%%%%%%%%%%%%%

In Table~\ref{EQM} we report the numerical values for the Euclidean constituent quark masses, $M^E_f$, obtained with the different combinations of interaction and vertex 
functions that enter the DSE kernel~\eqref{QuarkDSE}. As an illustration of the mass functions, we plot them for four flavors and in the chiral limit in Fig.~\ref{Massf} 
for the case of solving the DSE with the BC +T  vertex~\eqref{fullvertex} and the infrared-finite QC interaction~\eqref{QC}. 
The functional behavior is similar for other vertex ans\"atze, however the dressed mass $M(p^2)$ tends to reach its perturbative limit faster in the RL truncation than 
with the BC or BC + T vertices as a function of $p^2$. The common and more important feature is the mass function's fast rise and inflection point in the range,
1~GeV$^2\lesssim p^2 \lesssim 10$~GeV$^2$, which can be traced back to the lack of positive-definite spectral function of the quark propagator and thus 
confinement~\cite{Bashir:2012fs,Cloet:2013jya}.

It is clear from Table~\ref{EQM} and Fig.~\ref{Massf} that  DCSB plays a substantial role {\em even\/} for the charm quark since it is responsible for nearly 40\% 
of its constituent mass. Therefore, a careful treatment of heavy-light mesons ought to take  this feature into consideration and abandon a constant-mass propagator
for the charm~\cite{Rojas:2014aka,Mojica:2017tvh,ElBennich:2010ha,ElBennich:2011py,El-Bennich:2016bno,El-Bennich:2017brb}.

%%%%%%%%%%%%%%%%%%%%%%%%%%%%%%%%%%%%%%%%%%%%%%%%%%%%%%%%%%%%%%%%%%%%%%%%%%%%%%%%%%%%%%%%%%%%%
%
\begin{figure}[t!]
\vspace*{-3mm}
\hspace*{-3mm}
   \includegraphics[scale=0.6]{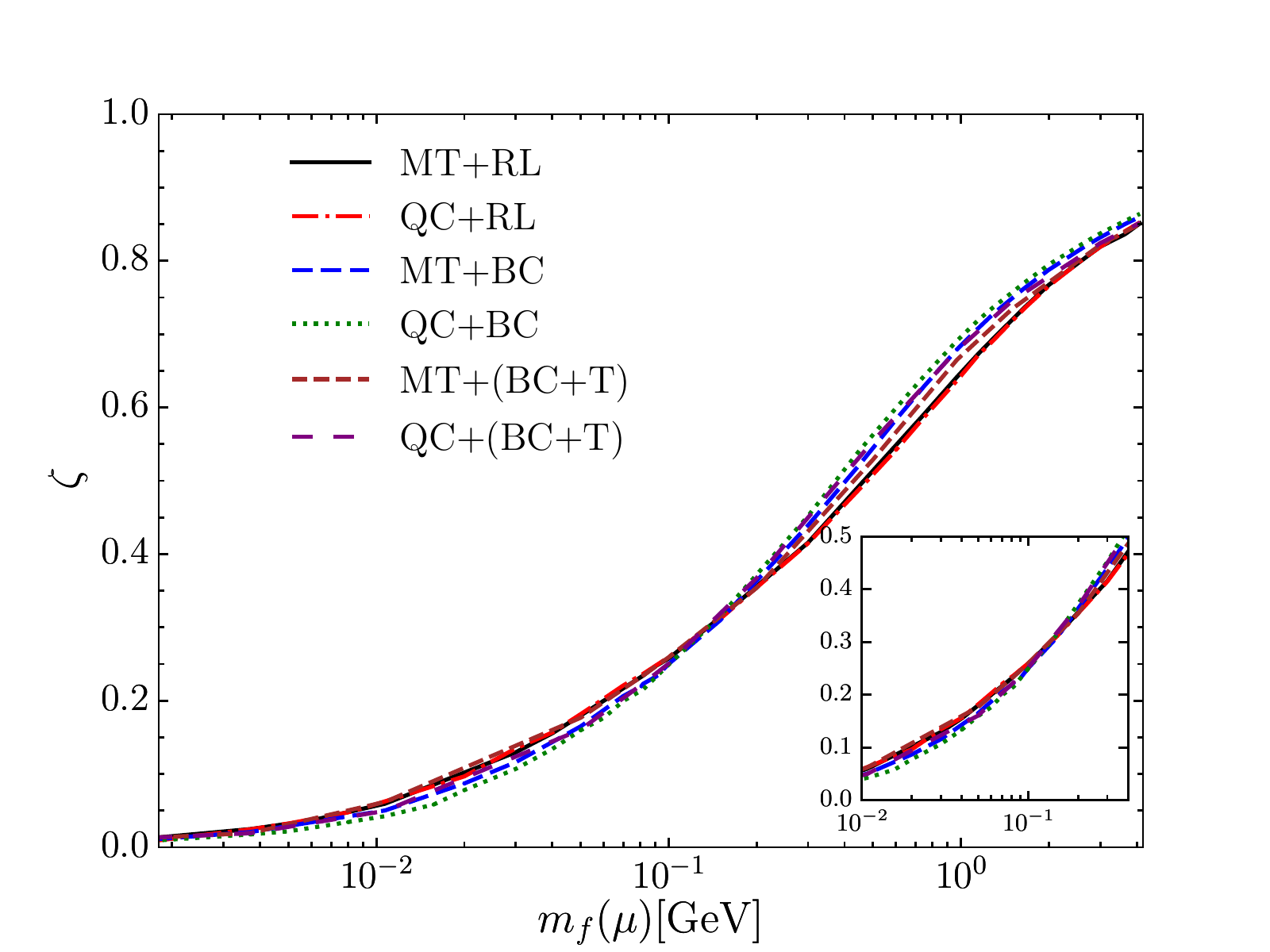}
\caption{The renormalization-point invariant ratio $\zeta$ as a function of $m_f(\mu)$ defined in Eq.~\eqref{zetadef}. Models and parameters are as in Table~\ref{Table:Results}. }
\label{zeta}
\end{figure}

%%%%%%%%%%%%%%%%%%%%%%%%%%%%%%%%%%%%%%%%%%%%%%%%%%%%%%%%%%%%%%%%%%%%%%%%%%%%%%%%%%%%%%%%%%%%%

\begin{table}[h!]
\caption{Flavor dependence of the renormalization-point invariant ratio $\zeta$~\eqref{zetadef}. Models and parameters are as in Table~\ref{Table:Results}. }
\def\arraystretch{1.5}
\centering\vspace*{1mm}
\begin{tabular*}%{|c|c|c|c|c|c|c|}\hline
{\hsize}
{
c@{\extracolsep{0ptplus1fil}}
c@{\extracolsep{0ptplus1fil}}
c@{\extracolsep{0ptplus1fil}}
c@{\extracolsep{0ptplus1fil}}
c@{\extracolsep{0ptplus1fil}}
}
\hline\hline
$f$ & $u,d$ & $s$ & $c$&$b$\\
\hline\hline
$\zeta^{\text{MT+RL}}$ & 0.025& 0.234&0.642&0.851\\
$\zeta^{\text{QC+RL}}$ & 0.025& 0.237&0.638&0.852\\  
$\zeta^{\text{MT+BC}}$ & 0.021& 0.215&0.679&0.860\\
$\zeta^{\text{QC+BC}}$ & 0.018& 0.216&0.691&0.864\\
$\zeta^{\text{MT+(BC+T)}}$ & 0.019& 0.235&0.665&0.850\\
$\zeta^{\text{QC+(BC+T)}}$ & 0.019& 0.224&0.678&0.852\\\hline\hline
\end{tabular*}
\label{zeta_f}
\end{table}

%%%%%%%%%%%%%%%%%%%%%%%%%%%%%%%%%%%%%%%%%%%%%%%%%%%%%%%%%%%%%%%%%%%%%%%%%%%%%%%%%%%%%%%%%%%%%

The core results of this study are summarized in Table~\ref{zeta_f} and Figure~\ref{zeta}, where we depict the evolution of the ratio $\zeta$ as function of the renormalized 
current-quark mass for the quark-gluon interaction models listed in Table~\ref{Table:Results}. We recover the well-known results for the RL truncation that the constituent-quark 
mass in case of a light quark is roughly 98\% due to DCSB, whereas for the $b$-quark it is merely about 15\%. Strikingly, it can be inferred from Table~\ref{zeta_f} that this 
observation is nearly independent of the interaction and vertex ansatz used in the DSE and this is true for all flavors.

One reads from Fig.~\ref{zeta} that $\zeta$ experiences a rapid rise in the range, 0.1~GeV $\lesssim m_f(\mu)  \lesssim1$~GeV, that is in the mass region between $m_s (\mu)$ 
and $m_c (\mu)$. Around $m(\mu) \approx 220$~MeV, an inflection point is followed by a continuing  and later attenuated increase of $\zeta$ towards its limiting value. In the RL 
truncations this increase beyond 0.5~GeV is slightly lower, indicating an enhanced DCSB contribution for heavier quarks compared to the vertices in Eqs.~\eqref{BallChiu} and
\eqref{fullvertex}.

As mentioned earlier, we treat the DSE kernel in Eq.~\eqref{STIkernel} separately for the sole reason the lattice-QCD simulations for the ghost and gluon propagators were performed 
with dynamical light quarks which range from 20 to 50 MeV~\cite{Ayala:2012pb} at $\mu = 2$~GeV. For consistency, we choose  $m_{u,d}(2~\mathrm{GeV} ) = 40$~MeV evolved 
to $\mu= 4.3$~GeV which is where the conditions \eqref{EQ:Amu_ren} and \eqref{massmu_ren} are imposed.  

As a consequence, for the light quarks in Table~\ref{zeta_f_lattice} the CSB contribution to the constituent quark mass is more importante when compared to the $\zeta$ values in 
Table~\ref{zeta_f} --- for light quarks one starts out with $\zeta_{u,d} = 0.253$, see Fig.~\ref{zetalattice}. In other words, the value of the ratio of CSB to the sum of CSB and DCSB 
is 0.25.  Since the Euclidean constituent quark mass  $M_{u,d}^E$ is about 36\% lighter with the lattice-generated interaction and the STI vertex~\eqref{STIvertex}, it is clear that important 
tensor structures responsible for DCSB were left out. The complete structure will be discussed elsewhere~\cite{Ahmed2018} and produces DCSB in amounts comparable to that 
exhibited in Table~\ref{EQM}. 

Nevertheless, the chiral symmetry breaking strength of this minimal vertex ansatz, which does not rely on any model interaction, produces realistic constituent-quark masses for 
current quark masses beyond $\sim 100$~MeV.  Therefore, the increase of $\zeta$ in Fig.~\ref{zetalattice} is initially nearly linear on logarithmic scale with again a slight inflection point 
at about $m(\mu) \approx 400$~MeV, whereas at  $m(\mu) \approx 220$~MeV CSB and DCSB appear to be of equal importance. The latter observation departs somewhat from what is 
seen in Fig.~\ref{zeta} where  equal contributions occur at $m(\mu) \approx 400$~MeV.

%%%%%%%%%%%%%%%%%%%%%%%%%%%%%%%%%%%%%%%%%%%%%%%%%%%%%%%%%%%%%%%%%%%%%%%%%%%%%%%%%%%%%%%%%%

\begin{table}[t!]
\caption{Euclidean quark masses, $M^E_f$, and renormalization-point invariant  ratio $\zeta$~\eqref{zetadef} obtained from the implementation of the integral kernel in Eq.~\eqref{STIkernel}.
For the solution of the quark DSE we make use of $\alpha_s=0.45$ and $\mu=4.3$~GeV. We also employ $m_{u,d}(\mu)=38.9$ MeV, $m_{s}(\mu)=82.1$ MeV, $m_{c}(\mu)=1.304$ GeV and 
$m_{b}(\mu)=4.697$~GeV,  which generate  $M_{u,d}(0)=0.272$~GeV, $M_{s}(0)=0.347$~GeV,  $M_{c}(0)=1.635$~GeV and $M_{b}(0)=4.874$~GeV, respectively.}
\def\arraystretch{1.5}
\centering\vspace*{1mm}
\begin{tabular*}
{\hsize}
{
c@{\extracolsep{0ptplus1fil}}
c@{\extracolsep{0ptplus1fil}}
c@{\extracolsep{0ptplus1fil}}
c@{\extracolsep{0ptplus1fil}}
c@{\extracolsep{0ptplus1fil}}
}
\hline\hline
$f$ & $u,d$ & $s$ & $c$&$b$\\
\hline\hline
$(M^E_f)^{\text{Latt+STI}}$ & 0.257& 0.324&1.547&4.671\\
$\zeta^{\text{Latt+STI}}$ & 0.253& 0.355&0.773&0.943\\\hline\hline
\end{tabular*}
\label{zeta_f_lattice}
\end{table}

%%%%%%%%%%%%%%%%%%%%%%%%%%%%%%%%%%%%%%%%%%%%%%%%%%%%%%%%%%%%%%%%%%%%%%%%%%%%%%%%%%%%%%%%%%%%%

%%%%%%%%%%%%%%%%%%%%%%%%%%%%%%%%%%%%%%%%%%%%%%%%%%%%%%%%%%%%%%%%%%%%%%%%%%%%%%%%%%%%%%%%%%%
%%%%%%%%%%%%%%%%%%%%%%%%%%%%%%%%%%%%%%%%%%%%%%%%%%%%%%%%%%%%%%%%%%%%%%%%%%%%%%%%%%%%%%%%%%%

\section{Final Remarks}

In this work we have investigated for which range of current quark masses the balance of CSB and DCSB is comparable in dependence of: 
\begin{itemize}
\item \ a chosen nonperturbative gluon-interaction model,
\item \ an ansatz for the dressed quark-gluon vertex,
\item \ a DSE kernel based on a minimal vertex from STIs and lattice-gluon and -ghost propagators.
\end{itemize}

As we have seen, this occurs somewhere midway between the strange and charm mass and is fairly independent of the ingredients in the quark-gap equation.
While the gauge-dependent quark-mass function, $M(p^2)$, reaches its perturbative limit faster in the RL models considered here than with 
gauging-technique vertices, the ratio $\zeta$ proves to be largely independent of the details of the integral kernel in the quark DSE. 

This, however is true when the combination of vertex and gluon dressings produces an interaction strength and functional form congruent with that required by hadron 
phenomenology, i.e. which reproduces the experimental hadron mass spectrum, weak decay constants and electromagnetic form factors.

We went a step further and analyzed the DCSB of a nonperturbative quark-gluon vertex model whose ``longitudinal" components constitute a ghost-improved 
Ball-Chiu vertex~\cite{Rojas:2013tza,Aguilar:2010cn}, whereas the transverse vertex consists of relevant tensor structures derived from two transverse 
STIs~\cite{He:2009sj}. We deliberately neglect contributions from the quark-ghost scattering kernel, as their effect is of minor order than that of the transverse 
vertex. Solving the DSE with kernel of Eq.~\eqref{STIkernel} we find massive solutions with considerable DCSB, attenuated for the light and strange quarks
and similar to that of phenomenological interactions for the heavy quarks. 

This latter DSE kernel does not satisfy the requirements for a sound description of  hadron properties, yet crucial improvements are underway~\cite{Ahmed2018} 
and the only inputs are gluon- and ghost-dressing functions for which numerical solutions of DSEs or from lattice QCD are readily available.  For the present purposes, 
this minimal STI vertex suffices to verify that CSB and DCSB are of roughly the same order at a mass scale of about 220~MeV, i.e. between the strange- and charm-quark 
mass.

%%%%%%%%%%%%%%%%%%%%%%%%%%%%%%%%%%%%%%%%%%%%%%%%%%%%%%%%%%%%%%%%%%%%%%%%%%%%%%%%%%%

\acknowledgements
 B.E. appreciated a stimulating discussion on chiral symmetry breaking with Yu-Xin Liu at Peking University. The authors are grateful to Gast\~ao Krein for 
 insightful comments on the manuscript and to Jos\'e Rodr\'iguez-Quintero for providing the Pad\'e approximations of the gluon- and ghost-dressing functions. 
 This work was partially  supported by CNPq grant nos.~168240/2017-3 (F.E.S) and 307485/2017-0 (B.E.), CAPES grant no 1811288 Edital 086/2013 (F.E.S) 
 and FAPESP grant nos.~2016/03154-7 (B.E.) and 2015/21550-4 (C.C).

%%%%%%%%%%%%%%%%%%%%%%%%%%%%%%%%%%%%%%%%%%%%%%%%%%%%%%%%%%%%%%%%%%%%%%%%%%%%%%%%%%%

\end{document}